\documentclass[aps,prd,reprint,groupedaddress,nofootinbib,floatfix]{revtex4-2}
\usepackage{color}
\usepackage{parskip}
\usepackage{amsmath,amssymb,amsfonts,bm}
\usepackage{graphicx}
\usepackage{float}
\def\Phib{\overline{\Phi}}
\def\deltab{\overline{\delta}}
\def\Tc{{\cal T}}

\begin{document}

\title{Fermions, quantum gravity and holography in two dimensions}

\author{Muhammad Asaduzzaman}
\email[]{masaduzzaman@uiowa.edu}
\affiliation{University of Iowa, Department of Physics and Astronomy, Iowa City, 52242, Iowa, USA}

\author{Simon Catterall}
\email[]{smcatterall@gmail.com}
\affiliation{Syracuse University, Department of Physics, Syracuse, 13244, New York, USA}

\author{Abhishek Samlodia}
\email[]{asamlodia@gmail.com}
\affiliation{Syracuse University, Department of Physics, Syracuse, 13244, New York, USA}

\date{\today}

\begin{abstract}
    We study a model comprising $N$ flavors of
    K\"ahler Dirac fermion propagating on a triangulated two-dimensional disk which is constrained
    to have a negative average bulk curvature.
    Dirichlet
    boundary conditions are chosen for the fermions. Quantum fluctuations of the geometry 
    are included by summing over all possible triangulations consistent with these constraints.
    We show in the limit $N\to \infty$ that the partition function is dominated by
    a regular triangulation of two-dimensional hyperbolic space. We use strong coupling expansions
    and Monte Carlo simulation to show that in this limit 
    boundary correlators of the fermions have a power law dependence on boundary separation as one
    expects from holography. However, we argue that this behavior breaks down for any finite
    number of massive fields 
    in the thermodynamic limit and quantum fluctuations of the bulk
    geometry drive the theory into a non-holographic phase. In contrast, for massless
    fermions, we find evidence that the boundary is conformal even for finite $N$. 
    This is consistent with theoretical results in quantum Liouville theory.
\end{abstract}

\maketitle

\section{Introduction}\label{sec:intro}
In this paper, we study the effects of fermionic matter on the partition function for two-dimensional quantum gravity. In particular we are interested in questions of holography and
hence restrict our discrete geometries to triangulations with the topology of a disk.
For our fermionic matter, we use K\"{a}hler-Dirac fermions rather than Dirac fermions.

K\"ahler-Dirac (KD) fields afford a natural way to couple
lattice fermions to discrete geometry. In a continuum $D$-dimensional Euclidean space with metric $g$,
a KD field $\Phi$ is composed of the set of
all (Grassmann valued) \\
antisymmetric tensor fields ($p$-forms)
\begin{equation}
    \label{eq:KD_field}
    \Phi = (\phi,\phi_{\mu},\phi_{\mu\nu},\ldots,\phi_{\mu_1\mu_2\ldots\mu_D})
\end{equation}
with the corresponding action,
\begin{equation}
    S_{\rm KD}=\int \mathrm{d}^D x\,\sqrt{g}\;\;\Phib\left(d-d^\dagger+m\right)\Phi
\end{equation}
where $d$ denotes the exterior derivative and the only dependence of the fermion operator on the metric arises in the
definition of the adjoint operator $d^\dagger$. Notice that the square of the KD operator
$(d-d^\dagger)$ is just the Hodge Laplacian. In the flat space, one can form a matrix fermion $\Psi$ using these fields as coefficients in an expansion over the Clifford algebra
of Dirac gamma matrices 
\begin{align}
    \Psi=\phi I+\phi_{\mu}\gamma^{\mu}+\phi_{\mu\nu}\gamma^\mu\gamma^\nu+\ldots \nonumber\\
    +\phi_{\mu_1\mu_2\ldots\mu_D}\gamma^{\mu_1}\gamma^{\mu_2}\cdots\gamma^{\mu_D}
\end{align}
It is then straightforward to show that $\Psi$ satisfies the usual Dirac equation $(\gamma^\mu\partial_\mu+m)\Psi=0$
and hence the fermionic content of the flat space theory is equivalent to a set of $2^{D/2}$ degenerate
Dirac fermions corresponding to the columns of $\Psi$. However, in a curved space this equivalence to the Dirac equation is lost.

A key advantage of KD fermions over Dirac fermions is they may be coupled to gravity without 
introducing frames and spin
connections and may be discretized without inducing
fermion doubling. 
Indeed one merely maps the $p$-form fields to lattice fields defined on $p$-simplices in
a triangulation ($p$-cochains)  and replaces $d$ and $d^\dagger$ by boundary $\delta$ and co-boundary $\deltab$
operators respectively, which have a natural action on such $p$-cochains - see \cite{banks_1982,Rabin:1981qj} and \cite{simon_kdf_2023,Catterall:2022jky,Butt:2021brl} for details.

In our work, we have focused on studying
the back reaction induced on the gravitational theory as a consequence of the presence of
KD fermions and particularly its effect on the holographic properties of the system. We employ
both analytic arguments and Monte Carlo simulation to infer the phase diagram of the theory. 
We find,
for a large number of fields $N$ , the effect of the fermions is to suppress curvature
fluctuations. In the limit that $N\to\infty$, we find that the geometry becomes a
classical constant (negative) curvature space corresponding to a regular tessellation of two-dimensional hyperbolic space.

Furthermore, we show in this limit, that 
boundary-boundary correlation functions exhibit
a power law decay as expected for a holographic theory. Indeed, the dependence on bulk mass matches
well with continuum predictions \cite{Klebanov_1999}. However, we will argue
that this behavior does not survive if the number of fields is held fixed
in the thermodynamic limit -- the discrete geometries become disordered as the area of the
disk increases and the simple holographic behavior of the correlators is lost.

\section{Model}
The partition function for the model is written as,
\begin{align}
    Z&=\sum_{\Tc\in \mathrm{disk}}\int D\Phib\,D\Phi\, e^{\Phib\,K(\Tc)\Phi}e^{-S_g}\nonumber\\
    &=\sum_{\Tc\in \mathrm{disk}} \prod_{p=0}^D\,{\rm det}^{N/2}\left(\Box^{(p)}(\Tc)+m^2\right)\,e^{-S_g}\nonumber\\
    &=\sum_{\Tc\in \mathrm{disk}}e^{-S_{\rm eff}}\,e^{-S_g}\nonumber\\
    &{\rm where}\quad S_{\rm eff}=-\frac{N}{2}\sum_{p=0}^D {\rm Tr}\,\ln{\left(\Box^{(p)}(\Tc)+m^2\right)}
\end{align}
where we have included $N$ flavors of KD fermion and the two-dimensional bare
gravitational action is just given by a cosmological constant term $S_g=\kappa N_2$ where $N_2$
is the number of triangles. In practice, we consider triangulations of the disk $\Tc$
with a fixed number of
triangles so this term plays no role. The square of the KD operator yields
an operator which is the direct sum of Laplacians $\Box^{(p)}$ for each type $p$ of simplices (vertices, links, and triangles) and hence appear in the effective action arising from the fermion loops.
The sum over $\Tc$ includes all combinatorial triangulations with a boundary structure
that corresponds to the last layer of a finite $\{3,7\}$ tessellation of hyperbolic space \cite{Asad2020_prd}. That is,
we impose the discrete equivalent of asymptotic AdS boundary conditions on the geometry. As a consequence, the average bulk curvature is fixed to a negative constant while a Dirichlet
boundary condition is used for the fermions.

\section{Analytic analysis}
It is instructive to analyze the model first in the large mass limit $m\to\infty$. Furthermore, let us restrict our attention
to the contributions of the $p=0$ (node) sector of the fermion
operator whose matrix elements take the form
\begin{equation}
    M_{ij}=\Box^{(0)}_{ij}+\delta_{ij}q_i m^2=(1+m^2)q_i\delta_{ij}-C_{ij}
\end{equation}
where $q_i$ denotes the number of neighbors of the $i^{\rm th}$ node and $C_{ij}=1$ if
nodes $i$ and $j$ are neighbors. It is easy to factorize this in
the form $M=QAQ$ where the matrix $Q$ is diagonal with matrix elements $\sqrt{q_i}$
and $A$ takes the form
\begin{equation}
    A_{ij}=(1+m^2)\delta_{ij}-\frac{C_{ij}}{\sqrt{q_iq_j}}
\end{equation}
Thus 
\begin{equation}
   {\rm det}( M ) = e^{N_0\ln{\left(1+m^2\right)}}\left(\prod_{i}^{N_0}q_i\right) e^{\sum_{L}\frac{1}{L}\frac{1}{\left(1+m^2\right)^L}\Omega_L}
\end{equation}
where $\Omega_L$ is the number of closed loops of length $L$ that can be drawn on the
triangulation with each loop
weighted by the inverse of the product of the $q_i$ for each vertex in the loop. Taking the large $m$ limit
we find that the effective action for KD fermions 
contains the local term
\begin{equation}
    S_{\rm eff} = -\frac{N}{2}\sum_{i \in \text{bulk}}\text{ln}(q_i) 
\end{equation}
Using the usual
expression for the scalar (Regge) curvature associated with each node in a triangulation
$R_i=\frac{\pi}{q_i}\left(q_i-6\right)$
this can be rewritten 
\begin{equation}
    -\frac{N}{2}\text{ln}(q_i) = {\rm const}+ N\sum_{i \in \text{bulk}}\frac{1}{\pi^2}R_i^2+\cdots
\end{equation}
Thus the leading effect of the fermion backreaction in the large mass limit
is to induce a $R^2$ operator that suppresses local curvature fluctuations. For
large $N$ this conclusion
can be reinforced by performing a steepest descent evaluation of this
leading contribution by solving
\begin{equation}\frac{\partial S_{\rm eff}}{\partial q_i} = 0\quad{\rm with}\; \sum_i q_i = {\rm fixed}\end{equation}.
This generates a homogeneous solution with $q_i=7$ - this value being determined by the geometrical
boundary conditions - here the fact that the boundary nodes have a connectivity
determined by the final layer of a $\{3,7\}$ tessellation.

At $m=0$ one can find
another representation of the determinant of the node Laplacian in terms
of the number of
spanning trees in the corresponding graph. Thus as $N\to\infty$ we expect the partition
function to be dominated by the triangulation with the maximal number of spanning trees -
the regular tessellation with constant local curvature \cite{Boulatov:1986jd}. Thus for
large $N$ and for both small and large fermion mass we expect the partition function
to be dominated by regular tessellations of two dimensional hyperbolic space. We will
see that this expectation is indeed borne out in our simulations.

Let us now examine the expected behavior of fermion boundary-boundary correlation functions on such regular
tessellations. We will concentrate on the node correlators although
our conclusions will apply equally to all components of the fermion correlator.
This correlation function can be written
\begin{equation}
    <\phi_i\phi_j>=\left(\frac{-K+m}{-\Box+m^2}\right)_{ij}
\end{equation}
where $K=\delta-\deltab$. Focusing on the piece proportional to $m$ we see that it can be written
as a matrix element of the inverse lattice Laplacian. Thus
\begin{equation}
    <\phi_i\phi_j>\sim \frac{1}{m}\left(1-\frac{1}{m^2}\Box\right)^{-1}_{ij}
\end{equation}
For large mass we can expand the inverse operator in powers of $1/m^2$. Each successive
term connects nodes one further step apart in the lattice. The first non-zero
contribution to this correlator then arises from the power of $\Box$ that corresponds to the
shortest path on the lattice between the two boundary nodes - the lattice
geodesic. For a regular hyperbolic lattice
this path runs through the bulk and is of length $\ln{r}$ where $r$ is the boundary separation.
The correlator is then
\begin{equation}
    <\phi_i\phi_j>\sim \left(\frac{1}{m^2}\right)^{\ln{r}}\sim \frac{1}{r^{2\Delta}}\quad{\rm with}\;\Delta=\ln{m^2}
\end{equation}
Thus holographic behavior is expected for any value of the bulk mass at least as $m\to\infty$.
It is not hard to generalize this
argument to show that the strong coupling limit of {\it any} lattice spin or
gauge model
formulated on a
tessellation of hyperbolic space will exhibit
holographic behavior - see \cite{Asaduzzaman:2021bcw}. 
Using Monte Carlo simulation
we will see that in fact this behavior extends to all values of $m$ provided the geometry
is homogeneous as was seen earlier in \cite{Asaduzzaman:2021ufo,Asaduzzaman:2022ppu}.

According to the AdS/CFT correspondence, the relationship between the mass $m$ of a scalar field in the bulk of $AdS_{d+1}$ and the scaling dimension $\Delta$ of the boundary field operator is given by the following equation~\cite{Klebanov_1999},
\begin{equation}
    \Delta(\Delta - d) = (mL)^2
\end{equation}
where $d$ is the dimension of the boundary and $L$ the AdS curvature. Two solutions exist for this equation which are,
\begin{equation}
    \label{eq:klebanvo_witten_formula}
    \Delta_{\pm} = \frac{d}{2} \pm \sqrt{\frac{d^2}{4} + (mL)^2}
\end{equation}
Using appropriate boundary conditions, one can get back either one of the solutions~\cite{Klebanov_1999}. Dirichlet boundary conditions as used here
ensure the $\Delta_+$
solution is obtained. 

This continuum prediction holds only for a classical homogeneous hyperbolic/AdS
space. One of the goals of our study is to see whether this behavior survives
in the presence of bulk quantum gravity fluctuations. In our
model the magnitude of the latter
can be adjusted by dialing the number of fermions propagating on the geometry. We have
argued that in the limit $N\to\infty$, the  background geometry will reduce to a hyperbolic geometry with constant local curvature and this is borne
out by our simulations which are discussed in the next section.

\section{Monte Carlo simulations}

\begin{figure}
    \centering
    \includegraphics[scale=0.2]{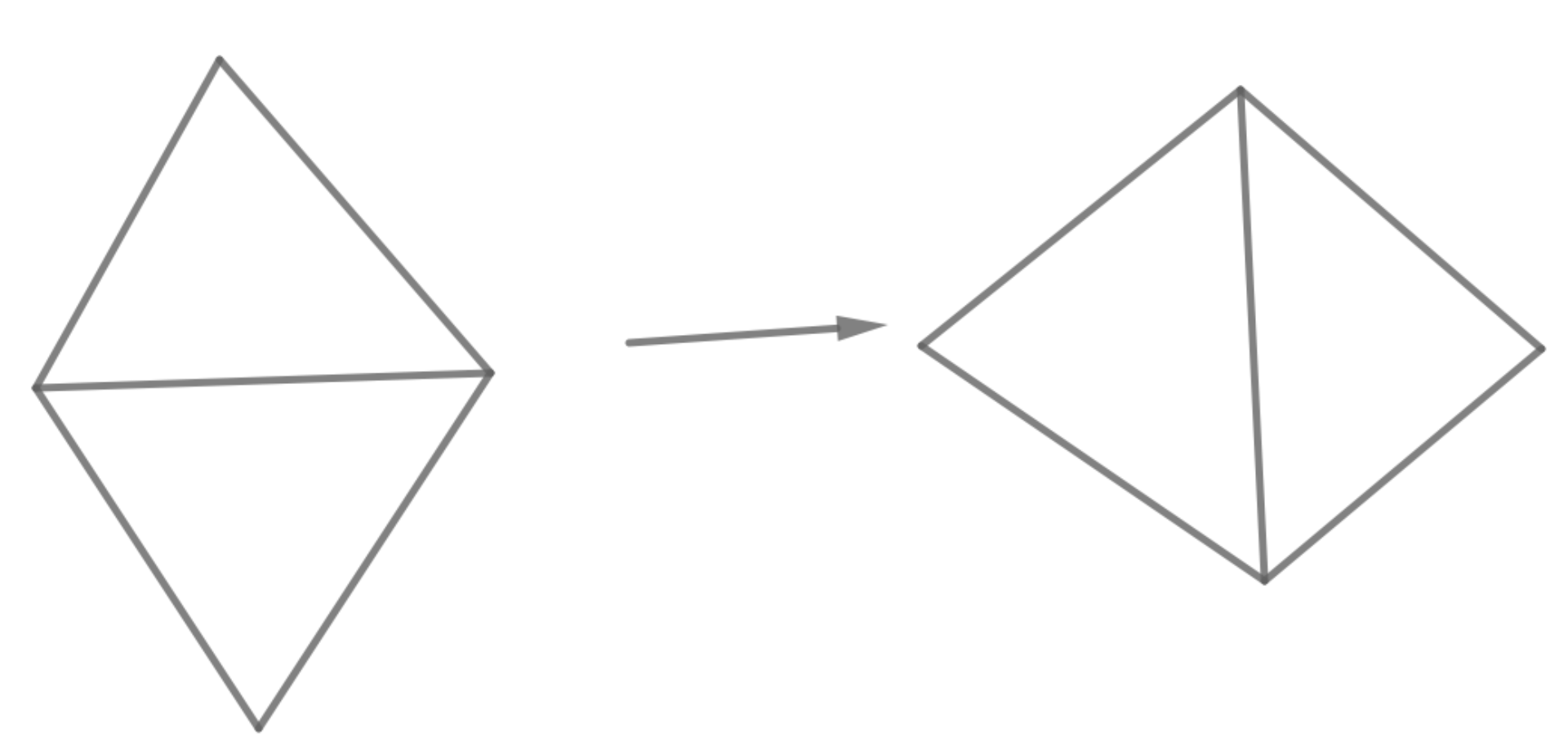}
    \caption{Link flip Move}
    \label{fig:linkflip}
\end{figure}

\begin{figure}
    \centering
    \includegraphics[scale=0.3]{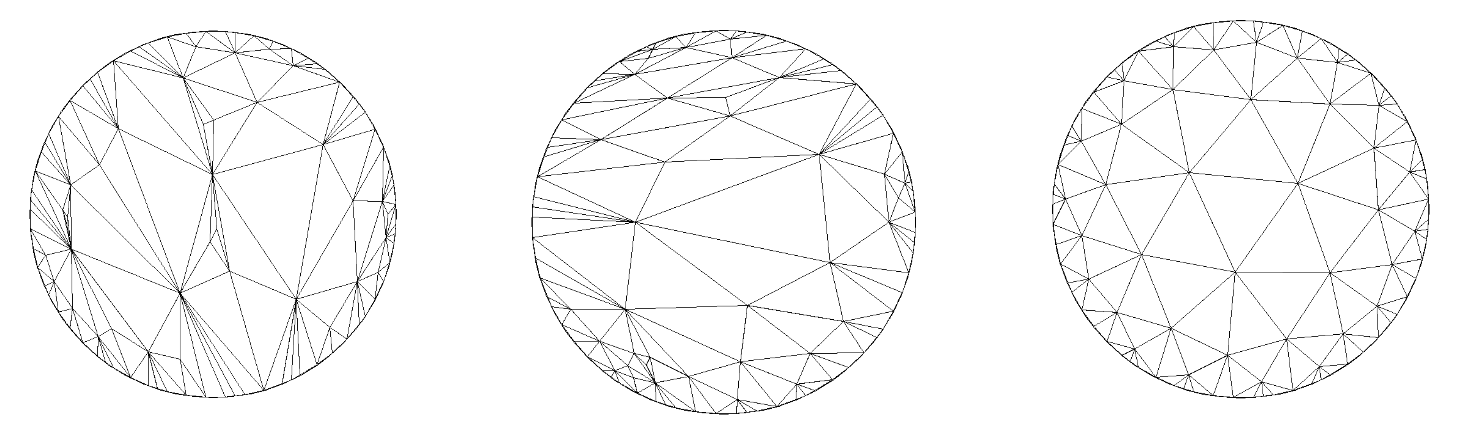}
    \caption{Typical geometries arising in Monte Carlo simulations. The disk on the right is a near perfect regular tessellation for large $N$ case.}
    \label{fig:geometries.}
\end{figure}

In order to do Monte Carlo simulations capable of
reproducing a path integral over bulk geometries we need a set of local
moves that change the triangulation $\Tc$.
A suitable
set of moves are called the Pachner moves and can be implemented in any
dimension - see reference-\cite{Catterall:1994sf} for a simple implementation of such moves
in $D$ dimensions.
In two dimensions one such move -- the so-called link flip move (see Figure~\ref{fig:linkflip}) -- is known to be ergodic in the space
of random triangulations of fixed area. By choosing to perform
this move at random
one can construct a Metropolis procedure that is capable of performing a random
walk in the space all such triangulations $\Tc$ weighted by the effective action. Figure~\ref{fig:geometries.} shows typical random geometries generated in Monte Carlo sampling, where near regular tessellations are obtained when $N$ is large.
The change in the effective action needed by the Metropolis algorithm
requires the evaluation of the change in the
fermion determinants
under such link flip moves. We have used the SuperLU package to compute the change in these
determinants~\cite{superlu99}. This is efficient but still scales like the cube of the system
volume which limits us to relatively small lattices.
In practice we have simulated a range of lattice areas up to $1162$ lattice sites using
variables numbers of KD fermions and using a set of masses that extend from
near massless $m=0.0001$ to $m=10.0$. Typical Monte Carlo ensembles for
a fixed set of parameters correspond to $1000$ configurations obtained from $20000$ Monte Carlo sweeps with a gap of $20$. 

\section{Simulations and Results}
\subsection{Bulk observables}
\begin{figure}
    \centering
    \includegraphics[scale=0.65]{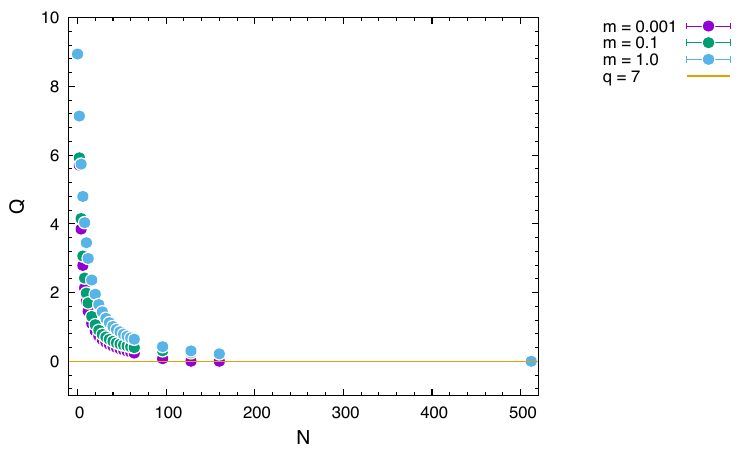}
    \caption{Curvature fluctuation $Q$ vs number $N$ of KD fermions with different bulk masses $m$ for a lattice consisting of $N_2=1162$ triangles. }
    \label{fig:q2_massless}
\end{figure}

\begin{figure}
    \centering
    \includegraphics[scale=0.6]{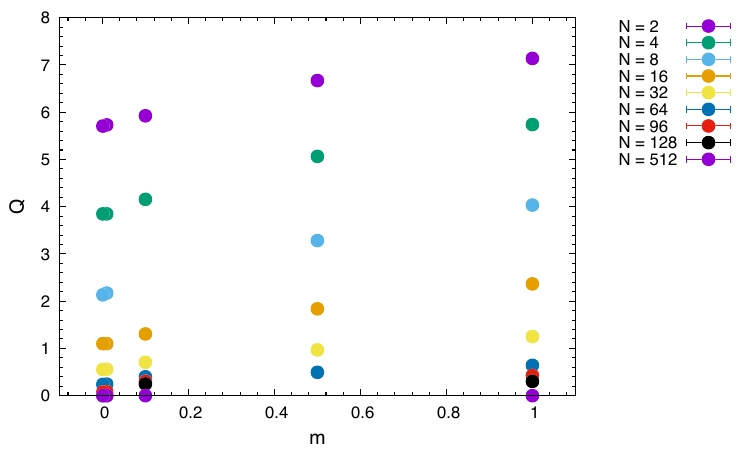}
    \caption{Q vs $m_{\text{bulk}}$ with different number of fields $N$ for a lattice with $N_2=1162$ triangles.}
    \label{fig:q2_vs_mbulk}
\end{figure}

In this subsection, we first investigate the bulk geometrical properties
of the triangulations.  
Figure~\ref{fig:q2_massless}, which plots the fluctuations in the
local curvature $Q = \langle(q-7)^2\rangle$ averaged
over the bulk nodes as a function of the number of
fields $N$, shows clearly that in the $N \to \infty$ limit, the geometry indeed
approaches the regular $\{3,7\}$ tessellation of hyperbolic space corresponding to
a continuum space with constant negative curvature.
However, it can be seen that the number of fields that are needed to freeze the lattice geometry to the
regular tessellation depends on both the mass $m$ and
the lattice area $N_2$. Figure~\ref{fig:q2_vs_mbulk} shows a plot of the bulk value of
$Q$ vs the bulk mass for $N=2-512$. Clearly one needs larger $N$ as $m$ increases to smooth
the lattice geometry
at fixed area. This makes sense; if one integrates out massive fields one would expect to generate
local $R^2$ operators as we saw for $m\to\infty$. However for finite mass there will
be $1/m^2$ corrections arising from the expansion over closed loops which reduce
the effective $R^2$ coupling. However this coupling will also
be proportional to the number of KD fields so the latter can be increased to compensate for
this - an effect which is visible in the plot.

\begin{figure}
    \centering
    \includegraphics[scale=0.65]{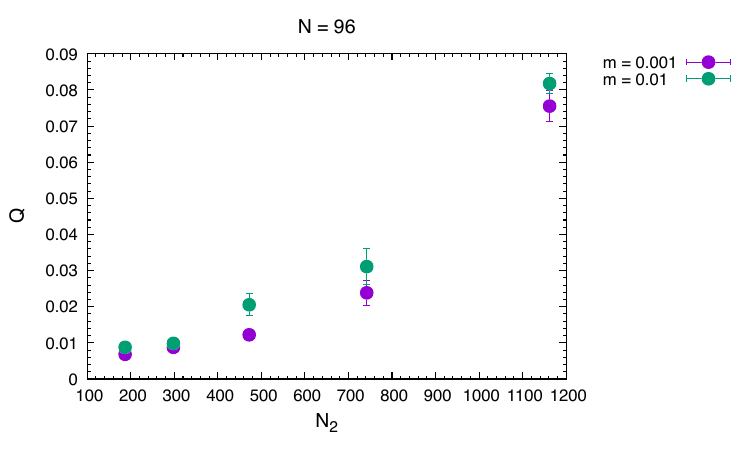}
    \caption{$Q$ vs area $N_2$ for fixed $N=96$ and different bulk masses.}
    \label{fig:q2_vs_vol_Nfixed}
\end{figure}

Indeed, since $R^2$ is an irrelevant operator in two dimensions one might expect its coupling to flow to zero in the infrared 
and hence to vanish in the thermodynamic limit rendering the bulk lattice
geometries disordered for any finite number of {\it massive} fermions.
In light of this it is interesting to ask what happens for vanishing mass as the
area is increased. Figure~\ref{fig:q2_vs_vol_Nfixed} shows
a plot of $Q$ vs the lattice area for several values of the bulk mass $m$ including
the small mass limit. It is clear
that the value of $Q$ indeed starts to grow as the area increases for any value of
the mass which
is consistent with our earlier argument. 

One can also understand this effect
via the following argument. Imagine the effect of a single link flip around the regular
tessellation. It is not hard to verify that this changes the action $\Delta S_{\rm 1 flip}=\alpha$
where $\alpha={\cal O}(1)$ and is {\it independent} of the area $N_2$. However there are many ways
to make such a fluctuation -- the link flip can occur on any bulk link. Thus the entropy
associated to such a flipped configuration increases logarithmically with the area. Putting these
two facts together one sees that the change in the free energy associated with a single
link flip $\Delta F=\alpha-\beta \ln{N_2}$  can be made arbitrarily negative by link flips for
large enough area. Thus one expects, even in the massless case,~\footnote{In practice $m=0.001$
gives a good approximation to the massless theory on our lattices. We have checked
that neither bulk or boundary observables change on further decrease of $m$} that link flips will
predominate for fixed $N$ in the large area limit - as we see in the data. Indeed, this
implies that the regular tessellation will \textit{not} dominate in the limit of
large areas {\it unless} one sends $N\to\infty$.

Thus our analysis of bulk local observables suggests that typical geometries become disordered
in the thermodynamic limit. We explore the consequences of this for holography in the next section.

\subsection{Boundary correlators}

\begin{figure}[H]
    \centering
    \includegraphics[scale=0.65]{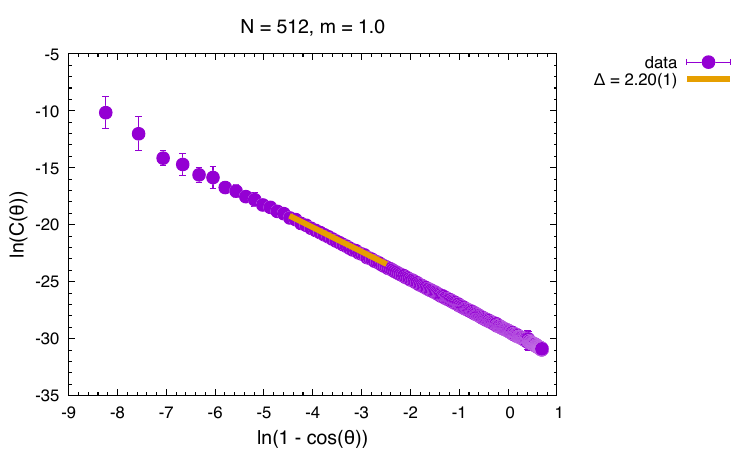}
    \caption{Correlation function for $N=512$ and $m=1.0$ using smoothing block size $b=4$ and a fitting window of size $20$.}
    \label{fig:m_1.0_N512_block4}
\end{figure}
The boundary correlation functions are computed as matrix elements of the inverse KD operator including the bulk mass term~\footnote{To compute the correlators we use open boundary conditions but include a very large boundary
mass for the fermions to drive the magnitude of the boundary field close to zero. Typically $m^2_{\partial D} = 1000.0$.}
\begin{equation}
    C(r)=\sum_{ij}\frac{1}{n_{ij}(r)}\delta_{\vert i-j\vert,r}\left(-\Box+m^2\right)^{-1}_{ij}
\end{equation}
where $\vert i-j\vert$ is the boundary distance (in units of the lattice spacing) between boundary
sites $i$ and $j$ and 
\begin{equation}
    n_{ij}(r)=\sum_{ij}\delta_{\vert i-j\vert,r}
\end{equation}

In practice we assess errors on our correlation functions by a jackknife procedure. Typically
we have used $N_{\rm jack}=200$ samples to assess statistical errors.
The boundary correlators for smaller $N$ suffer from large fluctuations as the 
boundary geodesic distance increases. These 
fluctuations become more significant for larger bulk mass.

To handle this and remove the leading discretization effects
we have also performed a smoothing operation on the correlation function before attempting power law
fits. This entails averaging the correlators at any distance
over a block of a certain size centered on that distance. We have checked that our results are robust to the size
of this parameter.
Once this is done we fit the resultant correlator to a power
law to extract $\Delta$.  

\begin{figure}[H]
    \centering
    \includegraphics[scale=0.65]{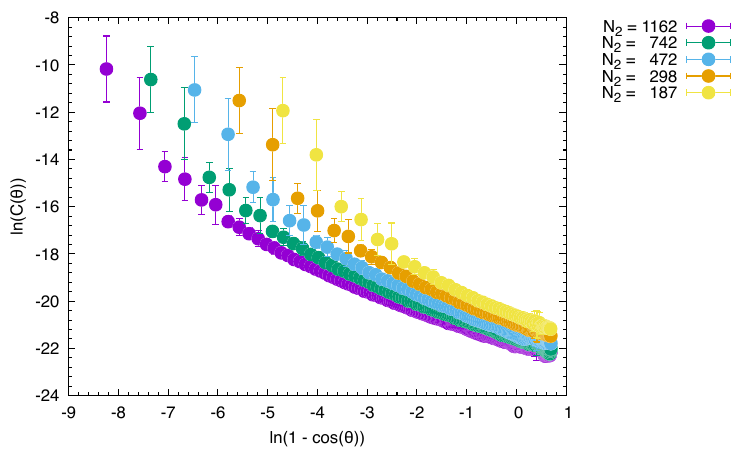}
    \caption{Correlation function for $N=2$ and $m=1.0$ with smoothing  block size $b=4$ for different number of simplices.}
    \label{fig:m_1.0_N2_smoothing_b4}
\end{figure}

An example of this procedure can be seen in Figure~\ref{fig:m_1.0_N512_block4} which shows
the smoothed correlation function and fit for $N=512$ and $m=1.0$. In this plot we have chosen to fit the data in a particular window of length 20 lattice spacings. We plot the data
versus $\ln{(1-\cos{\theta})}$ where $\theta=\pi\frac{r}{r_{\rm max}}$ and
$r_{\rm max}$ is the maximum boundary geodesic distance (half of the total length of the boundary). This takes into
account the leading finite size effects. For $r_{\rm max}\to\infty$ this
is just $2\ln{r}$. For large $N$
the results are somewhat insensitive to the choice of window. But this is no longer
true as $N$ is reduced. For example Figure~\ref{fig:m_1.0_N2_smoothing_b4} shows a correlator for $m=1.0$ at $N=2$ for several different values of the total area. In this case the log-log plot shows a pronounced curvature and a strong finite size effect and it is far less clear which fitting window should be chosen.

\begin{figure}[H]
    \centering
    \includegraphics[scale=0.65]{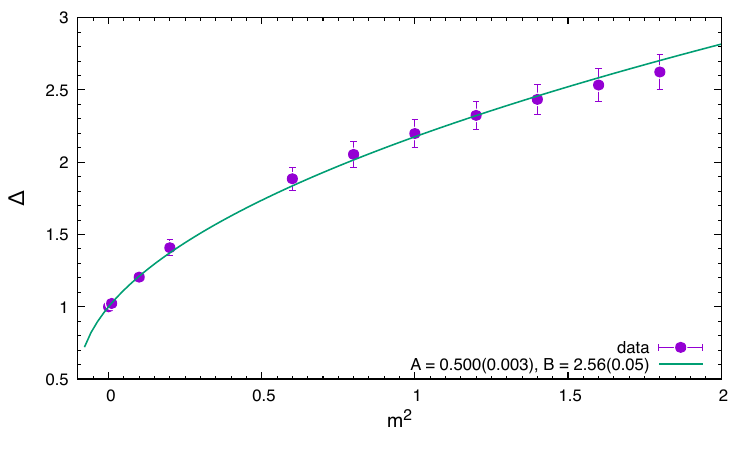}
    \caption{$\Delta$ vs $m^2$ at $N=512$ and $N_2=1162$. The
    $\chi^2$ per degree of freedom for this fit is 0.16}
    \label{fig:delta_vs_mbulk_N512}
\end{figure}
To establish robust values for
any extracted $\Delta$ we need to determine what fitting window to use. To do
this we have determined an effective running $\Delta$ at each distance $r$ and look for
a robust fitting window where consistent values of $\Delta$ are obtained independent of
the precise location of the window.
In more detail, we fix the fitting window size and move the fitting window progressively out to larger distances keeping both the block size $b$ and number of points $N_{\rm fit}$ fixed. For each choice of
fitting window, we fit a power law from $\ln{C(r)}$ vs $\ln{r}$ and extract
an estimate of the local
power law exponent $\Delta$. We then look for
a plateau in the value of $\Delta$ indicating a simple power law decay. We additionally check
that the resultant fits have an acceptable $\chi^2$ per degree of freedom (see the appendix
for a plot of $\chi^2$ vs the fitting window)
To test this method we first restrict ourselves to the case of $N=512$ where the lattices
are perfect regular $\{3,7\}$ tessellations of the hyperbolic plane for all masses and lattice
sizes we employed.
Figure~\ref{fig:delta_vs_mbulk_N512} plots $\Delta$ versus $m^2$ for $N=512$.
The fit we show corresponds to the continuum prediction given in eqn.~\ref{eq:klebanvo_witten_formula}. We fit the data with the form: $\Delta = A + \sqrt{A^2 + Bm^2}$. From the fit we find $A=0.500 \pm 0.003$ and $B=2.558 \pm 0.051$.
B is related to the radius of curvature $L = \sqrt{B/2}=1.131 \pm 0.022$. This can be compared with that expected for a $\{p,q\}$
tessellation of hyperbolic space given by
\begin{equation}
    \frac{1}{L}=2\cosh^{-1}{\left(\frac{\cos{(\pi/p)}}{\sin{(\pi/q)}}\right)}
\end{equation}
where $p=3$ and $q=7$ for a triangulation. The formula yields $L=0.917$ which differs by ${\cal O}(10)\%$ from the theoretical prediction which we attribute
to finite size effects~\footnote{The factor of $\sqrt{2}$ connecting $B$ to
$L$ arises from a dual lattice parameter that is needed to construct the correct kinetic
operator for the scalar field on a tessellation \cite{Asad2020_prd},\cite{Brower:2019kyh},\cite{brower2022hyperbolic} }.

\begin{figure}[H]
    \centering
    \includegraphics[scale=0.65]{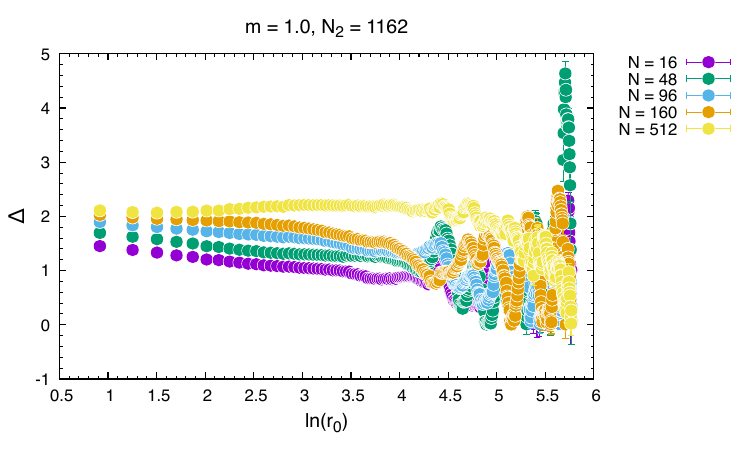}
    \caption{$\Delta$ vs location of fitting window for $N_2=1162$ and $m=1.0$ ($N_{\rm fit}=20$)}
    \label{fig:delta_lnr2_m_1.0_20}
\end{figure}

Let us now turn to the situation at finite $N$. 
Figure~\ref{fig:delta_lnr2_m_1.0_20} shows the value of $\Delta$ extracted
from simulations at $m=1.0$, area $N_2=1162$
for several values of $N$. We fix the fitting window to $20$ with $r_0$ the starting
point of the fit and the block size to $b=4$.
It should be clear that only for $N=512$ do we see a reliable plateau
while for smaller $N$ the effective scaling dimension $\Delta$ runs with distance. We interpret this as a breakdown
in holography. In appendix C we show that this conclusion does not depend on the size of
the fitting window $N_{\rm fit}$.

\begin{figure}[H]
    \centering
    \includegraphics[scale=0.65]{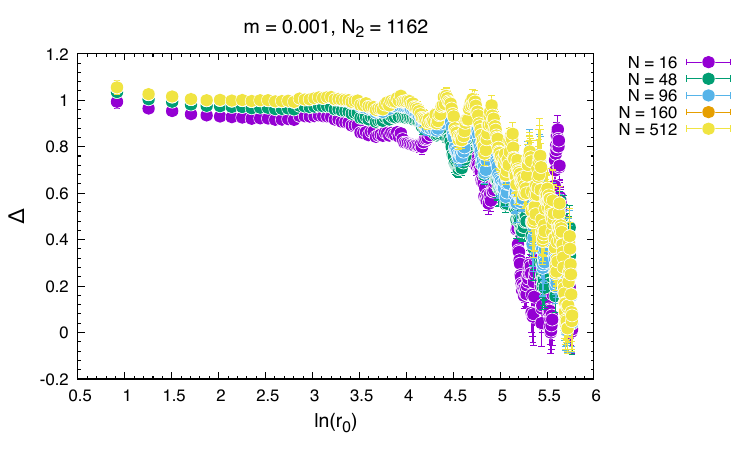}
    \caption{ $\Delta$ vs location of fitting window for $N_2=1162$ and $m=0.001$
    ($N_{\rm fit}=20$)}
    \label{fig:delta_lnr2_m_0.001_20}
\end{figure}

Figure~\ref{fig:delta_lnr2_m_0.001_20} shows a similar plot for $m=0.001$.
In this case a plateau is visible even for
small values of $N$~\footnote{In appendix~\ref{sec:appxB} we show a typical correlator in the small mass limit which reinforces the notion that one can obtain a good power law fit over a wide
range of fitting windows.}
Thus we conclude that the boundary theory remains conformal for massless fermions
even in the presence of bulk disorder although the
scaling dimension of the field appears to receive quantum gravity corrections. 
We discuss the possible reasons for this in our conclusions. In appendix~\ref{sec:appxC}
we show the roubustness of the results for $\Delta$ on the size of the fitting window by choosing
$N_{\rm fit} = 12$.

\section{Summary and Conclusion}
We have shown that the back reaction of KD fermions on a two-dimensional quantum geometry with the topology of a disk is such as to
suppress local curvature fluctuations. Indeed, for $N\to\infty$ limit, we find the lattice geometries approach that of a regular tessellation of hyperbolic space - in the limit a classical space with constant negative curvature. Furthermore, in this
limit, the boundary correlation functions exhibit a holographic property falling off as
powers of the boundary distance. Remarkably the dependence of this power on bulk fermion
mass
matches quite closely the continuum prediction. 

However, if the number of fields is held fixed, we have provided evidence that
fluctuations of the geometry reemerge in the thermodynamic limit both for massive
and massless fields. For massive fermions the boundary correlators, while the correlation is relatively long-ranged, no longer fit a simple power law.
We tentatively conclude that in this
regime the model
no longer exhibits a holographic phase.
This is not too surprising --  as we have shown, integrating out massive fields produces local operators such as $R^2$ which are irrelevant at long distances. The path integral
is then dominated by metrics that differ drastically from classical hyperbolic space.

The case of
massless fermions is more subtle -- in this case, non-local operators can arise in the
gravity action. For example the Polyakov action can be generated~\cite{DISTLER1989509,Mertens:2020hbs}
\begin{equation}
    S_{\rm P}=\int d^2x\sqrt{g(x)}\int d^2y\sqrt{g(y)}\,R(x)\,\Box^{-1}(x-y)\,R(y)
\end{equation}
where $\Box$ is the covariant Laplacian and $R$ the Ricci scalar.
In the continuum we can fix the conformal gauge $g=e^{2b\phi}\hat{g}$  and obtain the well
known Liouville action \cite{DISTLER1989509}:
\begin{equation}
    S=\frac{1}{4\pi}\int d^2x\,\sqrt{\hat{g}}\left[\left(\partial\phi\right)^2+Q\hat{R}\phi\right]
\end{equation}
where $Q=b+b^{-1}$ is related to the Liouville central charge $c_L=1+6Q^2$. The entire
system is quantum conformal invariant if $c_M+c_L=26$ where 26 arises from
the ghosts needed to gauge fix diffeomorphism symmetry. Notice that while
KD fermions are equivalent to multiples of Dirac fermions in flat space this is
no longer true in the presence of gravity where they behave like ghosts.
Indeed, for KD fermions, the central
charge $c_M\to -\infty$ as $N\to\infty$ and we are within the regime of
applicability of Liouville theory with $Q$ large and $b$ small corresponding to weak gravity. 
In this limit, we expect
that conformal invariance is maintained with boundary correlation functions
still falling off as a power of the distance \cite{Mertens:2020hbs} even though
the bulk geometry no longer corresponds to a classical hyperbolic space. This conclusion agrees with
our simulations.

\begin{acknowledgments}
Numerical computations were performed at Syracuse University HTC Campus Grid under NSF award ACI-1341006. We acknowledge support from U.S.
Department of Energy grants DE-SC0019139 and DE-SC0009998. 
\end{acknowledgments}

\appendix
\section{Fits for $\Delta$}\label{appx:data}
Below we show the result of fitting $\Delta$ for various fermion masses for a large
number of fermions and an area $N_2=1162$.
\begin{table}[H]
    \centering
    \begin{tabular}{|c|c|c|c|}
    \hline
    $m^2$ & $\Delta$ & $\sigma$ & $\chi^2 /{\rm d.o.f}$\\
    \hline
    0.000001 & 0.99936 & 0.00265 & 0.2469 \\
    0.0001 & 0.99959 & 0.00264 & 0.24682 \\
    0.01 & 1.02274 & 0.002 & 0.24036 \\
    0.1 & 1.20491 & 0.00805 & 0.21221 \\
    0.2 & 1.4084 & 0.00834 & 0.35023 \\
    0.6 & 1.88571 & 0.00975 & 0.30089 \\
    0.8 & 2.05425 & 0.01059 & 0.28498 \\
    1.0 & 2.19882 & 0.01066 & 0.29197 \\
    1.2 & 2.32405 & 0.01059 & 0.30889 \\
    1.4 & 2.43486 & 0.011 & 0.32482 \\
    1.6 & 2.53427 & 0.0117 & 0.34371 \\
    1.8 & 2.62441 & 0.01274 & 0.36496 \\
    \hline
    \end{tabular}
    \caption{Data for $\Delta$ vs $m^2$ at $N = 512, N_2 = 1162$. $\sigma$ is the error in estimation $\Delta$}
    \label{tab:delta_vs_mbulk_N512}
\end{table}

\section{Massless Fermion Correlators\label{sec:appxB}}
\begin{figure}[H]
    \centering
    \includegraphics[scale=0.65]{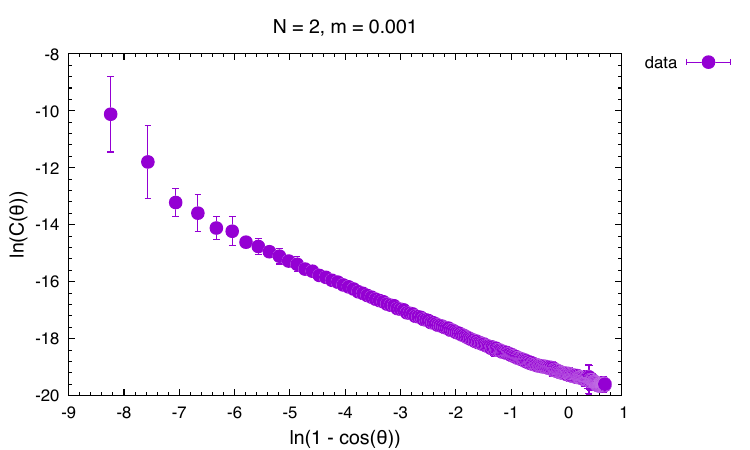}
    \caption{Boundary correlator with smoothing block size $b=4$f for $N=2$, $N_2=1162$ and $m=0.001$}
    \label{fig:m0.001_N2_b4_corr}
\end{figure}

Fig.~\ref{fig:m0.001_N2_b4_corr} shows a boundary correlator for small $N=2$ and in the (near) massless regime
$m=0.001$. A clear linear regime for large values of $\ln{(1-cos{(\theta)})}$ or $r$ indicates a good robust power law fit.
\section{Varying the fitting window\label{sec:appxC}}
\begin{figure}[H]
    \centering
    \includegraphics[scale=0.68]{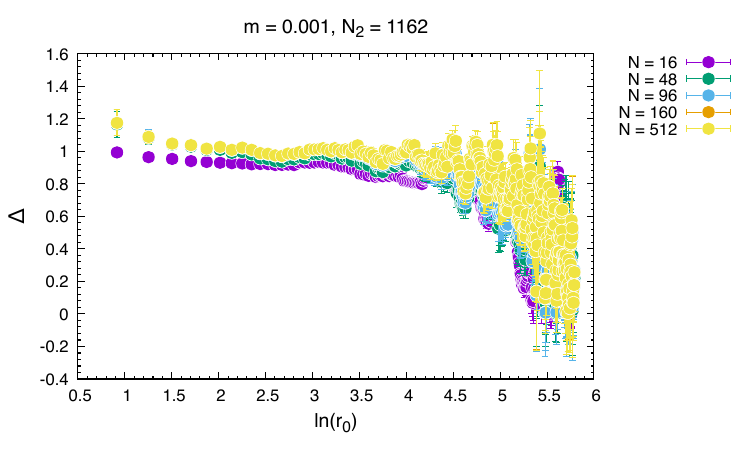}
    \caption{Scaling dimension $\Delta$ for fitting window $N_{\rm fit}=12$ with $N_2=1162$ and $m = 0.001$ for a variety of $N$}
    \label{fig:delta_lnr_m_0.001_12}
\end{figure}

\begin{figure}[H]
    \centering
    \includegraphics[scale=0.68]{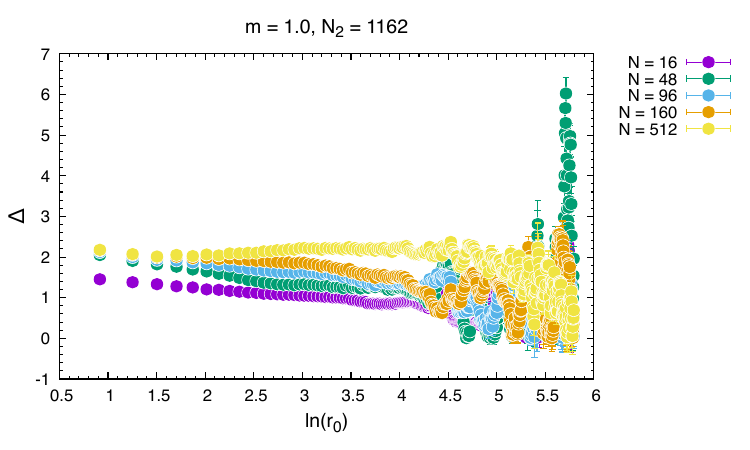}
    \caption{Scaling dimension $\Delta$ for fitting window $N_{\rm fit}=12$ with $N_2=1162$ and $m = 1.0$ for a variety of $N$}
    \label{fig:delta_lnr_m_1.0_12}
\end{figure}

\begin{figure}[H]
    \centering
    \includegraphics[scale=0.6]{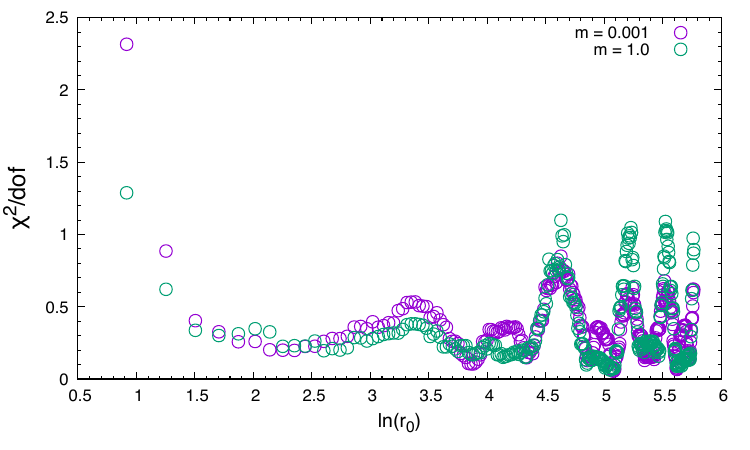}
    \caption{$\chi^2$ vs the fitting window for $N = 512$ and $m=1.0$}
    \label{fig:chisq_window_512}
\end{figure}

In the pictures below (Fig.~\ref{fig:delta_lnr_m_0.001_12},
\ref{fig:delta_lnr_m_1.0_12} and
\ref{fig:chisq_window_512}) we show the robustness of our results for
$\Delta$ on the size of the fitting window by choosing $N_{\rm fit}=12$ in contrast to the
choice $N_{\rm fit}=20$ used in the main text.

In  the following plot we show the $\chi^2$ of a typical correlator fit as a function
of the fitting window.


\bibliography{apstemplate}

\end{document}